\documentclass[aps,prl,twocolumn,10pt,showpacs,superscriptaddress,amsmath,amssymb,nobibnotes]{revtex4-1}

\usepackage{graphicx}
\usepackage{amsfonts}    
\usepackage{graphicx}   
\usepackage{verbatim}   
\usepackage{color}      
\usepackage{subfigure}  
\usepackage{hyperref}   
\usepackage{natbib}
\usepackage{booktabs}
\usepackage{threeparttable}
\usepackage{dcolumn}
\usepackage{multirow}
\usepackage{subfigure}

\begin{document}
	\title{Measuring Small Longitudinal Phase Shifts via Weak Measurement Amplification}
	\author{Kai Xu}\email{These authors contribute equally to this work}
	\affiliation{CAS Key Laboratory of Quantum Information, University of Science and Technology of China, Hefei 230026, China}
	\affiliation{CAS Center For Excellence in Quantum Information and Quantum Physics, University of Science and Technology of China, Hefei 230026, China}
	\author{Xiao-Min Hu}\email{These authors contribute equally to this work}
	\affiliation{CAS Key Laboratory of Quantum Information, University of Science and Technology of China, Hefei 230026, China}
	\affiliation{CAS Center For Excellence in Quantum Information and Quantum Physics, University of Science and Technology of China, Hefei 230026, China}
	\author{Chao Zhang}
	\author{Yun-Feng Huang}
	\author{Bi-Heng Liu}\email{bhliu@ustc.edu.cn}
	\author{Chuan-Feng Li}
	\author{Guang-Can Guo}
	\affiliation{CAS Key Laboratory of Quantum Information, University of Science and Technology of China, Hefei 230026, China}
	\affiliation{CAS Center For Excellence in Quantum Information and Quantum Physics, University of Science and Technology of China, Hefei 230026, China}
	\author{Meng-Jun Hu}\email{humj@baqis.ac.cn}
	\affiliation{Beijing Academy of Quantum Information Sciences, Beijing, 100193, China}
	\author{Yong-Sheng Zhang}\email{yshzhang@ustc.edu.cn}
	\affiliation{CAS Key Laboratory of Quantum Information, University of Science and Technology of China, Hefei 230026, China}
	\affiliation{CAS Center For Excellence in Quantum Information and Quantum Physics, University of Science and Technology of China, Hefei 230026, China}
	
	\date{\today}
	
	\begin{abstract}
		Weak measurement amplification, which is considered as a very promising scheme in precision measurement, has been applied to various small physical quantities estimation. Since many quantities can be converted to phase signal, it is thus interesting and important to consider measuring ultra-small longitudinal phase shifts by using weak measurement. Here, we propose and experimentally demonstrate a novel weak measurement amplification based ultra-small longitudinal phase estimation, which is suitable for polarization interferometry. We realize one order of magnitude amplification measurement of small phase signal directly introduced by Liquid Crystal Variable Retarder and show its robust to finite visibility of interference. Our results may find important applications in high-precision measurements, such as gravitational waves detection. 
		
	\end{abstract}
	
	\maketitle
	\section{Introduction}
	Weak measurement, which was first proposed by Aharonov, Albert and Vaidman \cite{AAV}, has attracted a lot of attention in the last decades \cite{rmp}. In the theoretical framework of weak measurements, the system with pre-selected state interacts weakly with the pointer first, and then followed by a post-selection on its state. When the interaction is weak enough such that only first-order approximation need to be considered, the so-called weak value of observable $\hat{A}$, defined as $\langle\hat{A}\rangle_{w}=\langle\psi_{i}|\hat{A}|\psi_{f}\rangle/\langle\psi_{i}|\psi_{f}\rangle$ with $|\psi_{i}\rangle$ and $|\psi_{f}\rangle$ are pre-selected state and post-selected state of the system respectively, emerges naturally in the framework of weak measurements \cite{AAV}. The weak value is generally complex, with its real part and imaginary part being obtained separately by performing measurement of non-commuting observables on the pointer \cite{Rotza}, and can be arbitrarily large when $|\psi_{i}\rangle$ and $|\psi_{f}\rangle$ are almost orthogonal. Although the weak value has been intensively investigated since its birth \cite{wv1,wv2,wv3,wv4,wv5,wv6,wv7}, the debate on its physical meaning continues \cite{de1,de11,de2,de21,de3,de31}. Regardless of these arguments, the method of weak measurement has been shown powerful in solving quantum paradox \cite{pa1,pa2,pa3,pa4,pa5}, reconstructing quantum state \cite{tom1, tom2, tom3, tom4,tom5,tom6,tom7,tom8,tom9,tom10}, amplifying small effects \cite{am1,am5,am8,am11,am2,am3,am4,am6,am7,am9,am10,am12,am13} and investigating foundations of quantum world \cite{f1,f2,f3,f4,f2018,f5,f6,f7,f8,f9}.  
	
	Among above applications, weak value amplification (WVA) is particularly intriguing and  has been rapidly developed in high precision measurements. In order to realize WVA, the tiny quantity to be measured need be converted into coupling coefficient of an von Neumann- type interaction Hamiltonian, which is small enough such that the condition of weak measurements is satisfied. The magnification of WVA is directly determined by the weak value of the system observable appearing in the interaction Hamiltonian. When the pre-selected state $|\psi_{i}\rangle$ and the post-selected state $|\psi_{f}\rangle$ of the system are properly chosen, the weak value can be arbitrarily large. However, the magnification is limited when all orders of evolution are taken into consideration \cite{lim1,lim2,lim3,lim4}.  While the potential application of the weak value in signal amplification was pointed out as early as 1990 \cite{sa}, it has drawn no particular attention until the first report on the observation of the spin Hall effect of light via WVA \cite{am1}. Since then, many kinds of signal measurements via WVA have been reported, such as geometric phase\cite{am13}, angular rotation \cite{an,an1}, 
	spin Hall effect\cite{am1,am5,am8,am11}, single-photon nonlinearity \cite{non1,non2}, frequency \cite{fre,am10}, etc. Meanwhile, WMA obtains the large weak value at the price of low detection probability by postselection, which may causes greater statistical error\cite{post1}.   Therefore, the controversy on whether or not WVA outperforms conventional measurements arises  \cite{co1,co2,co21,co22,co3,co4,co5}.
	Although not positive conclusions arrived by some theoretical researches, it becomes different when practical experiments are taken into consideration. There are some researches that have shown WMA has meaningful robustness against technical noise \cite{technoise1,technoise3,temnoise1,temnoise2}. Meanwhile, some technical advantages of WVA have been experimentally demonstrated \cite{tec2,tec3,detsat2,tec4}. Moreover, WVA-based proposals to achieve the Heisenberg limit  by using quantum resources such as squeezing\cite{co5} and entanglement\cite{enta1,enta2} are also explored, e.g., the entanglement-assisted WVA has been realized in optical system\cite{entae1,entae2}. Recently, Kim {\it et al} demonstrate a novel WMA scheme based on iterative interactions to achieve Heisenberg limit \cite{hei1}.

	Since many physical quantities can be converted into phase measurements, using WVA to realize ultra-small phase measurement, especially longitudinal phase, has been developed rapidly in recent years \cite{lon7,lon6,lon4,lon5}. However, the existing schemes are still not suitable for practical applications because of their severe requirements on the preparation of initial state of probe as perfect Gaussian distribution and detections on time or frequency domain \cite{lon7,lon1, lon3}. The direct amplification of the phase shift in optical interferometry with weak measurement has been experimentally studied by Li et al.\cite{lon4}. In that experiment, 
	the systerm state with initial phase shift is prepared before getting into interferometer, the weak measurement part is composed of a HWP and a sagnac-like interferometer,  and the phase shift can be directly measured  by scanning two oscillation patterns via a general polarization projection measurements device. Here we proposed a different scheme of ultra-small longitudinal phase amplification measurement within the framework of weak measurements. Compared with sagnac interferometer scheme in Ref.\cite{lon4}, it can be expanded to other interferometers like Michelson interferometer suggested in Ref.\cite{hu,hu2}. Surprisingly, no definite weak value occurs in our case and the magnification is nonlinear,  which makes it different from WVA. 
	In this Letter, we experimentally demonstrate this new scheme by measuring ultra-small longitudinal phase caused by the liquid crystal phase plate and realize one order of magnitude amplification.

	\section{Weak measurements amplification based phase measurement} 
	The key idea of weak measurements based ultra-small phase amplification (WMPA) is to transform the ultra-small longitudinal phase to be measured into a larger rotation along the latitude of Bloch sphere of the meter qubit, e.g., larger rotation of a photon's polarization \cite{hu}. To explicitly see how it works, consider a two-level system initially prepared in the state of superposition $|\psi_{i}\rangle_{S}=\alpha|0\rangle+\beta|1\rangle$ with $|\alpha|^{2}+|\beta|^{2}=1$. Contrary to most discussions of WVA in which continuous pointer is used \cite{AAV,rmp,am1,am2}, we adopt discrete pointer ,i.e., qubit \cite{tom1,f2,knee} prepared in the state of superposition $|\phi\rangle_{P}=\mu|\uparrow\rangle+\nu|\downarrow\rangle$ with $|\mu|^{2}+|\nu|^{2}=1$. The pointer can be another two-level system or the different degree of freedom of the same system. 
	We consider unitary control-rotation evolution of the system-pointer interaction
	\begin{equation}
	\hat{U}=|0\rangle\langle 0|\otimes\hat{I}+|1\rangle\langle 1|\otimes(|\uparrow\rangle\langle\uparrow|+e^{i\theta}|\downarrow\rangle\langle\downarrow|),
	\end{equation}  
	where $\theta$ is the ultra-small phase signal to be measured. After evolution of composite system, the post-selection is performed on the system that collapses it into state $|\psi_{f}\rangle_{S}=\gamma|0\rangle+\eta|1\rangle$ with $|\gamma|^{2}+|\eta|^{2}=1$. The state of the pointer, after the post-selection of the system, becomes (unnormalized)
	\begin{equation}
	\begin{split}
	|\tilde{\varphi}\rangle_{P}&=_{S}\langle\psi_{f}|\hat{U}|\psi_{i}\rangle_{S}\otimes|\phi\rangle_{P} \\
	&=\mu(\alpha\gamma+\beta\eta)|\uparrow\rangle+\nu(\alpha\gamma+\beta\eta e^{i\theta})|\downarrow\rangle
	\end{split}
	\end{equation}
	with the successful probability $P_{s}=\mathrm{Tr}[|\tilde{\varphi}\rangle_{P}\langle\tilde{\varphi}|]$ and $\alpha,\beta,\gamma,\eta$ are all taken to be real numbers without loss of generality. Since $\theta\ll 1$, $\alpha\gamma+\beta\eta e^{i\theta}=(\alpha\gamma+\beta\eta)e^{i\kappa}$ in the first order approximation with
	\begin{equation}
	\mathrm{tan}(\kappa)=\dfrac{\mathrm{sin}(\theta)}{\mathrm{cos}(\theta)+(\alpha\gamma)/(\beta\eta)}.
	\end{equation}
	Phase signal amplification is realized as the post-selected state is properly chosen such that $_{S}\langle\psi_{f}|\psi_{i}\rangle_{S}=\alpha\gamma+\beta\eta\rightarrow 0$. The normalized pointer state thus becomes
	\begin{equation}
	|\varphi\rangle_{P}=\mu|\uparrow\rangle+\nu e^{i\kappa}|\downarrow\rangle
	\end{equation}
	in the first order approximation. Analogous to micrometer, which transforms small displacement into larger rotation of circle, our protocol transforms ultra-small phase into larger rotation of pointer along latitude of Bloch sphere. The amplified phase information $\kappa$ can be easily extracted by performing proper basis measurement on the pointer. It is intriguing to note that the WMPA  seems to work even when $_{S}\langle\psi_{f}|\psi_{i}\rangle_{S}=0$ according to Eq. (3), in which case an infinitely large amplification can be realized. It is, however, not true because the relative phase signal reduces into global phase that cannot be extracted in that case according to Eq. (2).

	\begin{figure*}[tbp]
		\centering
		\includegraphics[scale=0.7]{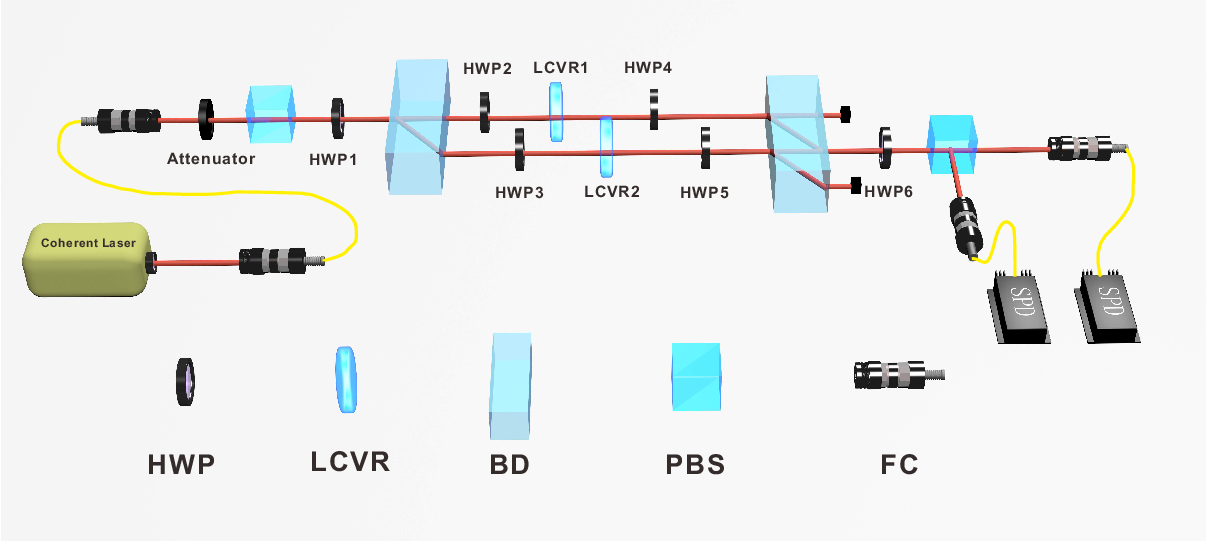}
		\caption{{\bf Experiment Setup}. An attenuated fiber-coupled laser beam is initially prepared in $|+\rangle$ polarization state by a polarizing beam splitter (PBS) and a half wave plate (HWP1) rotated at $22.5^{\circ}$. The initial state of the system and the pointer is prepared by a beam displacer (BD) and the HWP2 rotated at $22.5^{\circ}$, the HWP3 rotated at $67.5^{\circ}$, in which the system is the path degree of freedom and the pointer is the polarization degree of freedom. The ultra-small phase between horizontal and vertical polarization is produced by the Liquid Crystal Variable Retarder (LCVR1) and the LCVR2 without voltage added is used for phase compensation. The post-selection of the system is completed by the HWP3 rotated at $67.5^{\circ}-\delta$, the HWP4 rotated at $22.5^{\circ}-\delta$ and a BD with $\delta$ is an adjustable small angle to realize different magnification. The amplified phase signal, which encoded in the polarization state of post-selected photons, is extracted by polarization analyser consists of the HWP6 rotated at $22.5^{\circ}$ and a PBS. }
	\end{figure*}
	
	\section{Experiment realization} 
	In our experimental demonstration as shown in Fig. 1, we take the path state of photons as system and its polarization freedom of degree as pointer and perform ultra-small longitudinal phase measurement introduced by Liquid Crystal Variable Retarder (LCVR). We choose $\alpha,\beta,\mu,\nu=1/\sqrt{2}$ in our experiment such that the polarization of the post-selected photons is $(|H\rangle+e^{i\kappa}|V\rangle)/\sqrt{2}$ with $|H\rangle$ and $|V\rangle$ represent horizontal and vertical polarization respectively. The amplified phase $\kappa$ is extracted by performing measurement on the basis of $\lbrace |+\rangle,|-\rangle\rbrace$ with $|\pm\rangle=(|H\rangle\pm|V\rangle)/\sqrt{2}$ on the post-selected photons, which gives the expectation value of the observable $\hat{\sigma}_{x}\equiv |+\rangle\langle +|-|-\rangle\langle -|$ as $<\hat{\sigma}_{x}>=\mathrm{cos}(\kappa)$.
	
	The whole experimental setup consists of four parts i.e., initial state preparation, ultra-small phase signal $\theta$ collection, phase signal amplification via the post-selection and extraction of the amplified phase signal $\kappa$. The ultra-small phase $\theta$ is derived by substituting the measured $\kappa$ into the application formula Eq. (3), where $\alpha\gamma/\beta\eta$ is predetermined experimental parameter.
	
	As shown in Fig. 1, a single-mode fiber (SMF) coupled $808$ nm laser beam is emitted from the Coherent Laser (Mira Model 900-P). The laser beam has been attenuated before coupled into the fiber, which results in the final counting rate approximately $8\times 10^{5}/\mathrm{s}$. The light beam, which outputs SMF, passes a polarizing beamsplitter (PBS) and a half wave plate (HWP1) rotated at $22.5^{\circ}$ such that the polarization of photons is prepared in the $|+\rangle$ state. Preparation of the initial state of photons is completed by passing through a calcite beam displacer (BD) and two HWPs (HWP2 and HWP3) placed in the two paths separately. The BD is approximately $39.70$ mm long and photons with horizontal polarization $|H\rangle$ transmit it without change of its path while photons with vertical polarization $|V\rangle$ suffer a $4.21$ mm shift away from its original path. The HWP2 and HWP3 are rotated at $22.5^{\circ}$ and $-22.5^{\circ}$ respectively, which gives the initial state of photons as $(|0\rangle+|1\rangle)/\sqrt{2}\otimes|+\rangle$ with $|0\rangle$ represents the state of down path and $|1\rangle$ represents the state of up path. In fact, the initial state of the system i.e., path degree of freedom and the pointer i.e., polarization degree of freedom can be arbitrarily prepared via rotating HWP1 and HWP2, HWP3.

	The ultra-small longitudinal phase signal $\theta$ to be measured is produced by LCVR1 (Thorlabs LCC1411-B) placed in the up path. The LCVR causes phase shift between horizontal and vertical polarization state of photons when voltage is introduced by Liquid Crystal Controller (Thorlabs LCC25). Another LCVR placed in the down path without introducing voltage is used for phase compensation. The LCVRs fulfill the unitary control-rotation operation and the state of photons, after passing through the LCVRs, becomes
	\begin{equation}
	|\Psi\rangle_{SP}=\dfrac{1}{\sqrt{2}}[|0\rangle\otimes|+\rangle+|1\rangle\otimes(|H\rangle+e^{i\theta}|V\rangle)/\sqrt{2}].
	\end{equation}
	
	The amplification of the ultra-small phase signal $\theta$ is completed via HWP4, HWP5 and a BD, where the post-selected photons come out from the middle path of the BD toward to HWP6. To see explicitly how post-selection works, we can recast Eq. (5) as 
	\begin{equation}
	|\Psi\rangle_{SP}=\dfrac{1}{\sqrt{2}}[|H\rangle\otimes(|0\rangle+|1\rangle)/\sqrt{2}+|V\rangle\otimes(|0\rangle+e^{i\theta}|1\rangle)/\sqrt{2}],
	\end{equation}
	which implies the exchange of the path degree of freedom and the polarization degree of freedom of photons. Since the polarization degree of freedom of photons represents the system now, post-selection of the system can be readily realized by a HWP combined with a PBS. Suppose that the HWP is rotated at $22.5^{\circ}-\delta$ with $\delta$ is a small angle, then the post-selected state of photons coming out from the reflection port of the PBS is $|\psi_{f}\rangle=\mathrm{sin}(45^{\circ}-2\delta)|H\rangle-\mathrm{cos}(45^{\circ}-2\delta)|V\rangle$. After the post-selection, we exchange back the system and the pointer to the original degree of freedom by using a HWP rotated at $45^{\circ}$ in one of outgoing paths and a BD to recombine the light beam. The above process of post-selection can be equivalently realized via HWP4 rotated at $67.5^{\circ}-\delta$, HWP5 rotated at $22.5^{\circ}-\delta$ and a BD as shown in Fig. (1).
	
	\begin{figure}[tbh]
		\includegraphics[width=\linewidth]{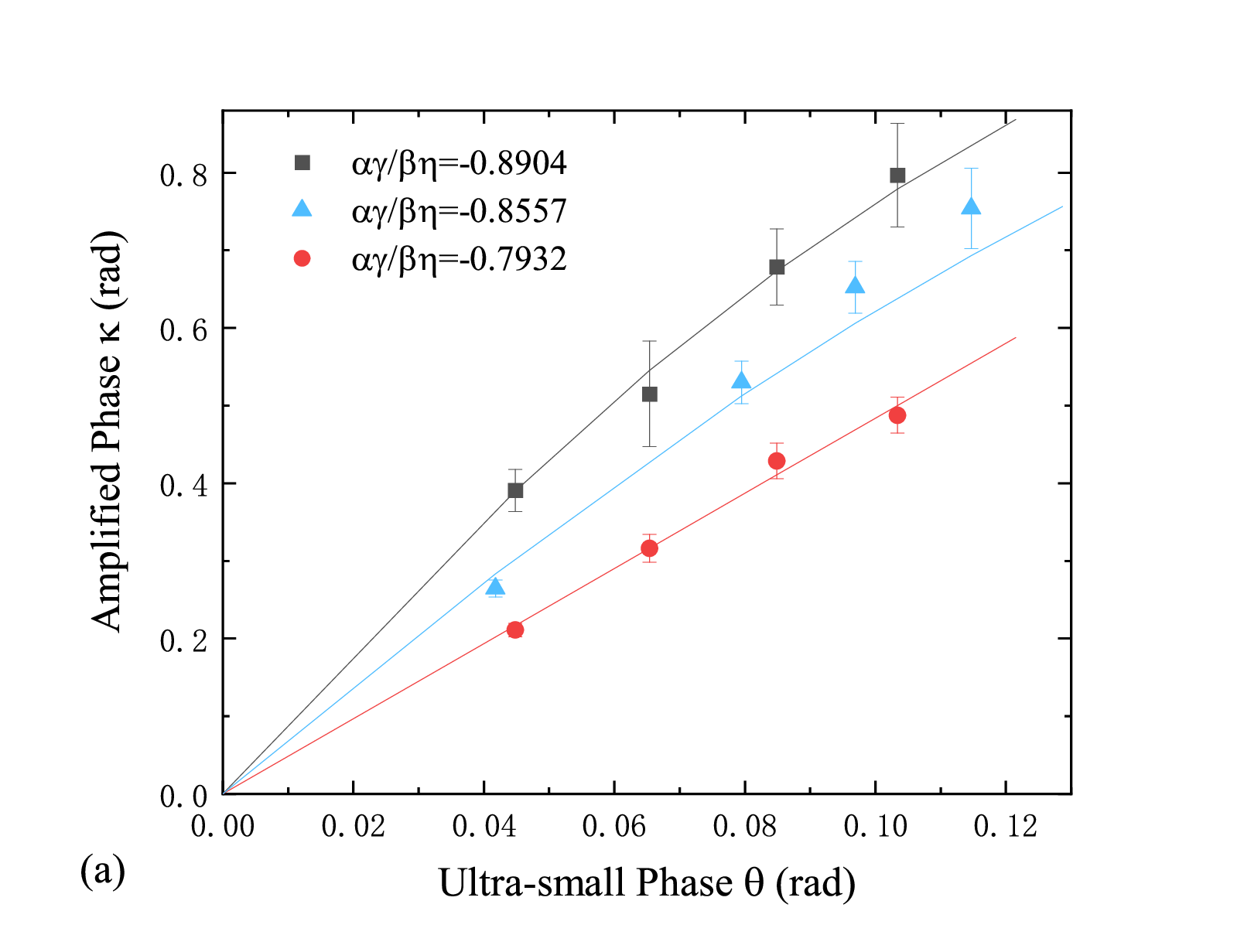}
		\includegraphics[width=\linewidth]{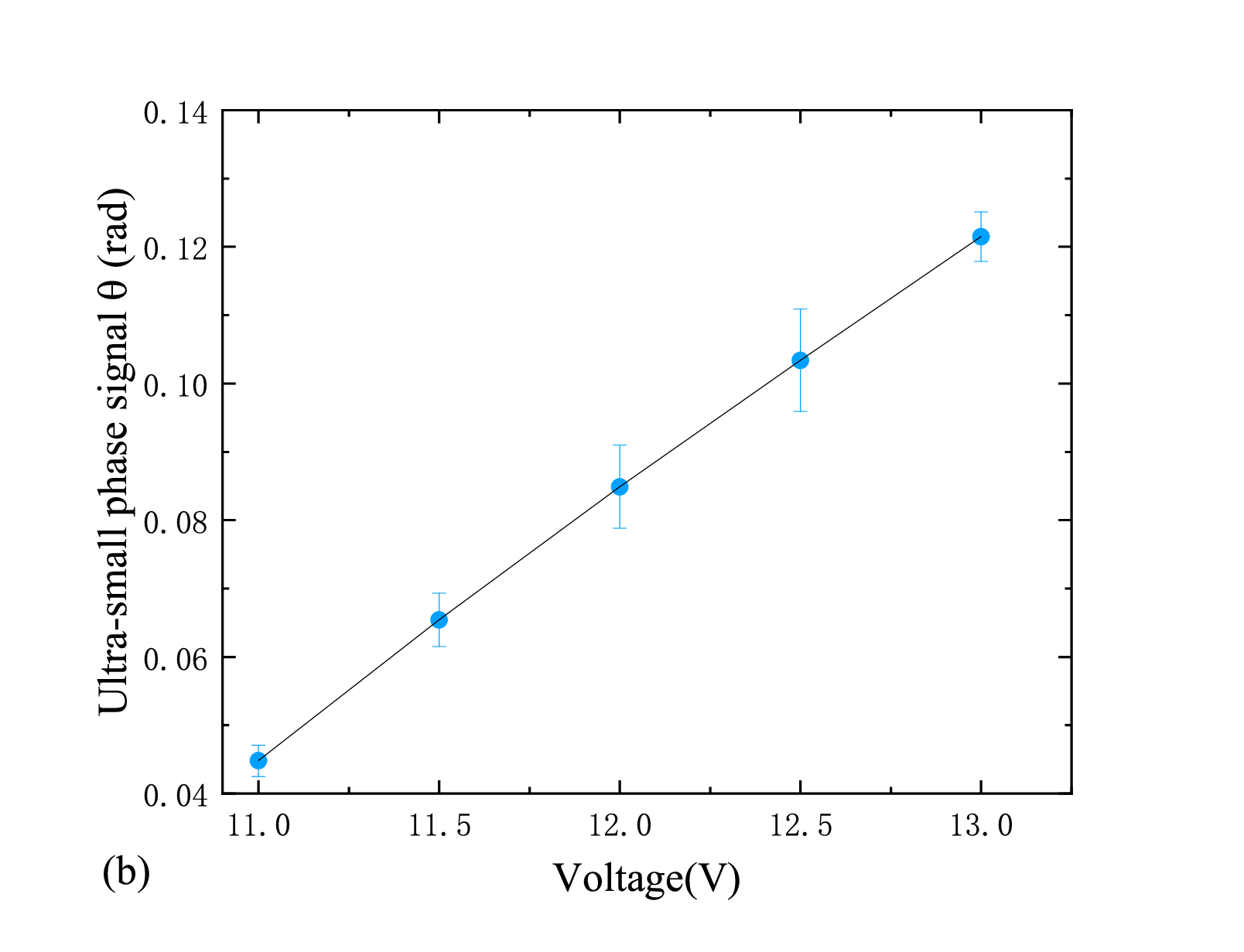}
		\caption{{\bf Experimental results of phase amplification and calibration}. (a) Measured amplified phase $\kappa$ vs ultra-small phase $\theta$. The real lines are theoretical predictions based on Eq. (3) with different colors representing different values of $(\alpha\gamma)/(\beta\eta)$. (b) Calibration curves of LCVR.}
		\label{fig:data2}
	\end{figure}

	The post-selected photons, which come from the middle port of the second BD, are in the polarization state $|\varphi\rangle_{P}=(|H\rangle+e^{i\kappa}|V\rangle)/\sqrt{2}$, where $\kappa$ is the amplified phase signal determined by Eq. (3) with $\alpha=\beta=1/\sqrt{2}$ and $\gamma=\mathrm{sin}(45^{\circ}-2\delta),\eta=-\mathrm{cos}(45^{\circ}-2\delta)$. The amplified phase $\kappa$ can be extracted by performing measurement on the basis of $\lbrace|+\rangle,|-\rangle\rbrace$ and calculating the expectation value of Pauli observable $\hat{\sigma}_{x}$, which is realized by the HWP6 rotated at $22.5^{\circ} $ ,a PBS and two avalanche photodiode single-photon detectors (SPD). Once the $\kappa$ is obtained, the ultra-small phase $\theta$ can be easily derived from the Eq. (3).
	
	\section{Results}
	Our experiment results are shown in Fig. 2.  Fig. 2(a) shows the relationship between amplified phase $\kappa$ and ultra-small phase $\theta$, in which real lines are theoretical predictions and dots are measured data. When phase signal to be measured is small enough, according to Eq. (3), the factor of amplification mainly determined by parameter $(\alpha\gamma)/(\beta\eta)$. Three different values of $(\alpha\gamma)/(\beta\eta)$ are considered in our experiment corresponding to about $3, 5$ and $10$ times magnification in the linear amplification region. For each case, four ultra-small phases chosen in the range of $0.03\mathrm{rad}-0.1\mathrm{rad}$, which are produced by LCVR1, are measured in advance. From Fig. 2(b), we can see the LCVR's phase is linearly related to the voltage.
	
	As an important experimental parameter, $(\alpha\gamma)/(\beta\eta)$ needs to be determined before the amplification measurement. This is done by measuring the successful probability of post-selection i.e., $p=|\langle\psi_{f}|\psi_{i}\rangle|^{2}=(\alpha\gamma+\beta\eta)^{2}$ without introducing any ultra-small phase. In the case of our experiment, $(\alpha\gamma)/(\beta\eta)=-\mathrm{tan}(45^{\circ}-2\delta)$ and $\mathrm{sin}(2\delta)=\sqrt{p}$, which gives 
	\begin{equation}
	\dfrac{\alpha\gamma}{\beta\eta}=\dfrac{\sqrt{p}-\sqrt{1-p}}{\sqrt{p}+\sqrt{1-p}}.
	\end{equation}
	Once parameter $(\alpha\gamma)/(\beta\eta)$ is settled, the voltage is added to LCVR to produce a ultra-small phase and the amplification measurement begins. The ultra-small phase $\theta$ is immediately estimated by conventional measurement method after amplification measurement, which is done by blocking the down path between BDs, replacing HWP4 with a HWP rotated at $67.5^{\circ}$ and rotating HWP6 to $45^{\circ}$. The visibility is about $0.999598$ so that the precision of phase estimation is about $0.04\mathrm{rad}$. Three different values of $(\alpha\gamma)/(\beta\eta)$ are obtained by adjusting the small angle $\delta$ and four ultra-small phases are measured within $10\mathrm{s}$ counting for each value. Considering the relevant statistical errors, system errors and imperfections of optical elements, our results meet well with theoretical predictions.
	
	As the key part of the experimental setup, the performance of the BD-type Mach-Zehnder interferometer directly determines the precision of phase estimation. The visibility of the interferometer in our experiment is about $0.9976$, which gives the precision of phase estimation about $0.098\mathrm{rad}$. In the amplification case, the ultimate precision of phase estimation should be divided by corresponding factor of amplification $h$, which implies that higher precision can be obtained compared to conventional Mach-Zehnder interferometry. The sensitivity of phase can also be significantly improved by weak measurements amplification if quantum noise limitation is not considered. The sensitivity in our amplification case is $\Delta\theta=\dfrac{\Delta\langle\hat{\sigma}_{x}\rangle}{h\mathrm{sin}(h\theta)}$ with $\Delta\langle\hat{\sigma}_{x}\rangle$ represents the uncertainty of $\langle\hat{\sigma}_{x}\rangle$, which implies $h$ times improvement even at the optimal point. When quantum noise limitation is considered, the ultimate sensitivity of weak measurements amplification cannot outperform the conventional measurements because of the large loss of photons. Fortunately, we need not worry about quantum noise too much in most practical experiment except for in super-sensitivity experiment such as gravitational wave detection and even in this kind of experiment, weak measurements amplification is able to approach the quantum noise limitation \cite{hu3}.
	
	\section{Discussion and Conclusion}
	Although we only experimentally demonstrate amplification of polarization-dependent longitudinal phase, the general phase amplification can be readily realized by using Michelson interferometer suggested in Ref.\cite{hu,hu2}. This realization indicates that WMPA is capable of measuring any ultra-small phase signal with higher precision and sensitivity than conventional interferometers in practice. 
	
	In conclusion, we have described and demonstrated a weak measurements amplification protocol i.e., WMPA that is capable of measuring any ultra-small longitudinal phase signal. The ultra-small phase introduced by LCVR is measured and one order of magnitude of amplification is realized. Larger amplification is possible if post-selected state is properly chosen. The WMPA would has higher precision and sensitivity than conventional interferometry if the quantum noise limitation is negligible, which is usually the case in practice. In addition, the precision of our scheme has potential to achieve the Heisenberg-limited precision scaling by using quantum resources such as squeezing\cite{co5}. Our results significantly broaden the area of applications of weak measurements and may play a crucial role in high precision measurements. 
	
	\section{Acknowledgment}
	This work is supported by the National Natural Science Foundation of China (No. 92065113, 11904357, 62075208 and 12174367), National Key Research and Development Program of China (No. 2021YFE0113100). Meng-Jun Hu is supported by Beijing Academy of Quantum Information Sciences.
	
	
	%

\end{document}